\documentclass[11pt]{article}

\newcommand{\reportversion}{v1.0}

\usepackage[margin=1in]{geometry}
\usepackage{setspace}
\usepackage{titlesec}

\usepackage{amsmath}
\usepackage{amssymb}
\usepackage{graphicx}
\usepackage{hyperref}
\usepackage{booktabs}
\usepackage{multirow}
\usepackage{float}
\usepackage{placeins}

\titleformat{\section}{\large}{\thesection}{1em}{}
\titleformat{\subsection}{\normalsize}{\thesubsection}{1em}{}

\titlespacing*{\section}{0pt}{1.5ex}{0.8ex}
\titlespacing*{\subsection}{0pt}{1.2ex}{0.6ex}

\hypersetup{
    colorlinks=true,
    linkcolor=blue,
    citecolor=blue,
    urlcolor=blue
}

\newcommand{\reportnumber}{BR-TR-2026-01}
\newcommand{\reporttitle}{FLUID: Slack-based Low-latency Delivery}
\newcommand{\reportauthors}{Michael Luby}
\newcommand{\reportaffiliation}{BitRipple, Inc.}
\newcommand{\reportdate}{May 2026}

\title{\reporttitle}

\author{
    \reportauthors\\
    \reportaffiliation\\
    Berkeley, CA
}

\date{
    \reportdate\\[0.5em]
    {\small \reportnumber\ (\reportversion)}
}

\begin{document}

\maketitle


\begin{abstract}

We introduce FLUID (Fountain LiqUId Delivery), a protocol that uses
fountain coding and receiver feedback for low-latency delivery of data
blocks over lossy networks. Idealized Automatic Repeat reQuest (ARQ)
protocols are bandwidth-optimal, but must deliver
every packet in a block and therefore can require additional rounds under
packet loss. FLUID uses a controlled amount of slack to relax this
all-packets requirement, allowing delivery to finish once enough encoded
packets have been received. This yields substantially tighter delivery
latency while remaining deterministically close to the ARQ bandwidth
optimum.

FLUID is controlled by a slack parameter $\epsilon$. Under the Loss-Product
Rule, delivery finishes once the product of packet loss fractions across
transmission rounds falls below $\epsilon$. Thus, FLUID can finish delivery
in a small number of rounds even when every round experiences packet loss,
while $\epsilon$ controls the gap between FLUID and bandwidth-optimal ARQ.

\end{abstract}

\newpage

\section{Introduction}

We introduce FLUID (Fountain LiqUId Delivery), a delivery protocol that uses
fountain coding to reliably deliver data blocks over lossy networks with low
latency and high delivery efficiency. FLUID transmits encoded packets that
each carry information about the entire data block.

Idealized Automatic Repeat reQuest (ARQ)~\cite{bertsekas1992data}
protocols based on retransmission, such as TCP~\cite{jacobson1988congestion},
provide the natural bandwidth-optimal reference point for single-path
reliable delivery. In the idealized ARQ model, every received packet is
source data, and each lost packet is retransmitted until it is received.
Thus, ARQ achieves delivery efficiency $1$ and is transmission-optimal for
the task of delivering every packet in the block. Its cost is latency:
because every packet must be delivered, packet loss can force additional
rounds of retransmission.

FLUID is parameterized by a slack parameter $\epsilon$, which controls how
much slack FLUID uses relative to this bandwidth-optimal ARQ reference point
in exchange for tighter delivery latency. Delivery latency is governed by the
Loss-Product Rule. Let $f_r$ denote the fraction of packets lost for a block
in transmission round $r$. Block delivery finishes in the earliest round
$\ell$ for which
\[
\prod_{r=1}^{\ell} f_r \le \epsilon.
\]
Thus, FLUID does not require any round to be loss-free. Instead, the loss
fractions multiply across rounds, and delivery finishes once the
loss-product falls below the slack parameter.

The same parameter $\epsilon$ also controls the deterministic delivery
efficiency gap from ARQ. For a block of $K$ source packets, FLUID uses a
block budget
\[
N = \left\lceil \frac{K}{1-\epsilon} \right\rceil.
\]
Thus, the delivery efficiency of FLUID is guaranteed to satisfy
\[
\frac{K}{N} \ge (1-\epsilon)\cdot \frac{K}{K+1},
\]
which approaches $1-\epsilon$ as $K$ grows. Under the comparison model used
in this paper, the ratio of FLUID transmissions per delivered source packet
to ARQ transmissions per delivered source packet is at most
\[
\frac{N}{K} \le \frac{1}{1-\epsilon} + \frac{1}{K},
\]
which approaches $\frac{1}{1-\epsilon}$ as $K$ grows.

This gives a precise latency-bandwidth tradeoff. ARQ remains the
bandwidth-optimal idealized baseline, but delivery depends on receiving
every packet. FLUID uses a deterministic amount of slack to finish delivery
after receiving enough encoded packets, yielding substantially tighter
delivery latency while remaining close to the ARQ bandwidth optimum.

This perspective also distinguishes FLUID from prior coded transport
approaches such as Network Coding Meets TCP~\cite{sundararajan2011network}
and Tetrys~\cite{lacan2012tetrys}. These designs use coding to improve loss recovery, but they do not provide
the same combination of guarantees: a simple rule that directly relates
delivery latency to the realized loss sequence, together with deterministic
control of the gap to bandwidth-optimal idealized ARQ. FLUID provides this
combination. The Loss-Product Rule determines the delivery round from the
observed loss fractions, while the slack parameter $\epsilon$ controls both
delivery efficiency and the transmission cost ratio relative to ARQ.

\section{Related Work: Reliable Delivery Protocols}

Existing reliable delivery protocols differ along three main dimensions:
how they respond to loss, how redundancy is introduced, and what feedback
is used to trigger additional transmissions. Retransmission-based protocols
recover specific missing packets. Rate-based FEC schemes add redundancy in
advance or according to estimated loss conditions. Coded and scheduling-based
transport protocols combine coding, rate control, or scheduling decisions
with feedback. These design choices affect delivery latency, delivery
efficiency, and transmission cost.

\paragraph{Retransmission-based protocols.}
TCP~\cite{rfc793} is the canonical retransmission-based reliable delivery
protocol, using a sliding window and packet-level retransmission without
coding. In the idealized Automatic Repeat reQuest (ARQ) abstraction used in
this paper, packets are retransmitted individually upon loss, and delivery
finishes when all source packets in the block have been received. This yields
delivery efficiency equal to $1$, since $K$ source packets are delivered as
$K$ useful packets.

Actual retransmission protocols introduce additional timing and signaling
effects. Practical TCP implementations use mechanisms such as selective
acknowledgments, retransmission timers, and modern time-based loss detection
to infer and repair losses~\cite{rfc2018,rfc6298,rfc8985}. A retransmission
may be triggered while the original packet or an earlier retransmission is
still in flight, causing the same data to be transmitted, and sometimes
received, more than once. Conversely, a retransmitted packet may itself be
lost, and detecting that loss may require an additional feedback cycle or
timeout before another retransmission is sent. These effects can increase
both latency and transmission cost relative to the bandwidth-optimal
idealized ARQ model. Even under the idealized model, however, missing
packets are recovered individually, so delivery may require multiple rounds
under packet loss, with finishing depending on delivery of each outstanding
packet.

\paragraph{Rate-based FEC schemes.}
Protocols such as Tetrys~\cite{lacan2012tetrys}, RTP/WebRTC FEC~\cite{rfc5109},
and QUIC-FEC variants~\cite{michel2019quicfec} introduce redundancy
proactively. Tetrys employs a sliding window with continuous repair packets
inserted at a fixed or adaptively tuned rate, while RTP/WebRTC FEC uses
fixed-size blocks with predetermined redundancy ratios, often combined with
retransmission for losses beyond FEC capability. QUIC-FEC variants combine
block or sliding-window FEC with ARQ. More generally, FECFRAME defines a
framework for applying FEC to packet flows over unreliable transport,
including real-time and streaming media applications~\cite{rfc6363}.

In these schemes, redundancy is set either statically or based on estimated
loss conditions. Recovery depends on whether the realized losses fall within
the provisioned redundancy or are subsequently repaired through additional
transmissions or retransmission. As a result, the number of rounds required
for delivery depends on the interaction between the loss realization and the
chosen redundancy, rather than being characterized by a simple rule tied
directly to the realized loss sequence.

\paragraph{Coded and scheduling-based transport protocols.}
A different class of approaches integrates coding or adaptive scheduling into
transport. Network Coding Meets TCP~\cite{sundararajan2011network} applies
random linear coding over a sliding window, tracking progress through degrees
of freedom (``seen'' packets). Recovery occurs when enough independent coded
packets have been received. While this approach can mask losses and maintain
throughput, it does not provide a simple delivery rule that directly
characterizes delivery latency.

S-Track~\cite{chu2023strack} adopts a packet-level scheduling approach,
selecting transmissions based on past delay and loss observations. Similarly,
coded multipath transport~\cite{poularakis2020multipath} uses block-based
coding across multiple paths, allocating traffic and redundancy based on
estimated path characteristics. These approaches rely on prediction,
optimization, or scheduling decisions; if path characteristics are
misestimated, the allocated redundancy may be insufficient, requiring
additional transmissions and increasing delivery latency.

\paragraph{Feedback mechanisms.}
A further distinction is the feedback used to trigger additional
transmissions. Retransmission-based protocols use identity-based feedback:
the sender must determine which data packets are missing, either directly or
indirectly, and retransmit those packets. This makes recovery sensitive to
out-of-order arrivals, stale feedback, lost retransmissions, and timeouts.
A retransmission may be unnecessary if the original packet is still in flight,
while a lost retransmission may not be detected until an additional feedback
cycle or timeout.

Rate-based FEC and scheduling-based schemes use feedback differently, but
still typically use it to manage repair rates, coding windows, degrees of
freedom, or scheduling decisions. Thus, feedback controls rate selection,
repair allocation, or packet identity decisions rather than directly giving a
complete block-local measure of how much work remains for delivery.

FLUID uses count-based feedback instead. Because fountain-encoded packets for
a block are interchangeable, the sender does not need to identify which
particular data packets were lost. The receiver reports how many encoded
packets have been received for the block together with the highest sequence
number received, allowing the sender to determine how many transmissions have
been reported lost. Each new feedback report supersedes the previous state for
the block, and each additional FLUID transmission is a newly generated encoded
packet rather than a retransmission of a specific packet. This makes the
feedback loop simpler and more reactive than identity-based retransmission.

\paragraph{Contrast with FLUID.}
Across prior approaches, redundancy is either implicit, fixed in advance, or
determined through rate selection and prediction. Feedback is typically used
to identify missing packets, track degrees of freedom, tune repair rates, or
guide scheduling decisions. Recovery is achieved through retransmission,
receipt of enough coded information, or scheduling decisions, but delivery is
not governed by a simple rule that directly relates latency to the realized
loss sequence together with deterministic control of delivery efficiency and
transmission cost relative to bandwidth-optimal ARQ.

FLUID differs in that delivery is governed by an explicit condition tied
directly to the realized loss sequence, while delivery efficiency and
transmission cost relative to bandwidth-optimal ARQ are controlled by
deterministic bounds. Rather than selecting a fixed repair rate or predicting
future loss, FLUID uses a block budget and extends transmission in response
to reported losses. Because fountain-encoded packets are interchangeable,
count-based feedback is sufficient to determine the amount of additional
transmission required for a block. This yields a simple rule relating
delivery latency to loss and a tunable tradeoff between delivery latency,
delivery efficiency, and transmission cost.

\section{FLUID Protocol Overview}

FLUID operates on a continuous stream of data. Source packets
arriving at the sender are grouped into blocks based on a short time
interval, and each block is encoded and delivered independently. This
block structure reflects the underlying object or timing structure of
the data stream, such as frames in a video application.

Each block has $K$ source packets that together represent the data to be delivered reliably, where $K$ can vary block to block. The sender generates encoded packets using a fountain erasure code, such as those described in \cite{luby2002lt,shokrollahi2009raptor, rfc6330}. The receiver can recover the original block once it has received $K$ encoded packets for the block.

The sender controls transmission using a path load parameter $\lambda \ge 1$, which defines the block budget for a block of data. For a block with $K$ source packets, the block budget is defined as
\[
N = \left\lceil \lambda \cdot K \right\rceil.
\]
Thus, $\lambda$ specifies the encoded packet budget for each block as a fraction of the number of source packets.

Initially, the number of encoded packets that may be transmitted is $N$. As transmission proceeds, this allowable number increases by the number of encoded packets reported lost to the sender. Thus, at any point in time, the total number of encoded packets that may be transmitted for the block is $N$ plus the cumulative number of reported losses.

Fountain codes make the FLUID protocol possible. In FLUID, the total number of encoded packets transmitted for a block is not fixed but depends on reported losses. Fountain code implementations enable the sender to generate additional encoded packets in response to reported losses and enable the receiver to recover the block from any $K$ received encoded packets.

The receiver monitors arriving packets and frequently transmits feedback reporting the number of encoded packets received for the block together with the highest packet sequence number received. From this information, the sender determines the cumulative number of encoded packets received and lost.

Transmission for a block terminates once the sender receives feedback indicating that the receiver has recovered the block or the sender block timer expires. Upon receiving this indication, the sender stops transmitting packets for the block.

FLUID is incorporated within the LT3~\cite{AggarwalLubyMinder2025Immersive} system, a network-layer packet delivery system.
The underlying protocol mechanisms are described in~\cite{luby2024fluid}.

\section{Loss-Product Analysis of FLUID and ARQ}

\subsection{Comparison Protocol Model Used for Analysis}

To enable a direct comparison between FLUID and ARQ, the analysis uses a
comparison protocol model that captures the causal behavior of packet loss
detection and response.

In this model, whenever a packet loss becomes inferable at the receiver,
the receiver immediately generates feedback indicating the loss. Feedback
messages are assumed to be reliably delivered to the sender after one
round-trip time (RTT) and are not themselves lost. Upon receiving feedback
indicating one or more losses for a block, the sender immediately transmits
one additional packet for each reported loss. For ARQ, each additional packet
is a retransmission of the corresponding lost source packet. For FLUID, each
additional packet is a newly generated encoded packet for the block.

This comparison protocol model reflects the structure of FLUID feedback,
which is count-based: the receiver reports the total number of packets
received for the block together with the highest sequence number received.
Each feedback message supersedes previous feedback for the block. From this
information, the sender determines how many encoded packets have been
received and how many have been reported lost, and can transmit new encoded
packets in response to the reported losses.

In contrast, the comparison protocol model is optimistic for ARQ. ARQ
protocols rely on identity-based feedback, often based on the highest
consecutively received packet. Missing packets may be inferred indirectly,
loss detection may be delayed by out-of-order arrivals, and retransmissions
may be triggered only after additional feedback cycles or timeouts.
Retransmitted packets may themselves be lost, causing further delay. For
the purposes of analysis, we idealize ARQ so that it follows the same
immediate loss detection and response behavior. Thus, the comparison protocol
model captures the feedback structure of FLUID but is optimistic for an
actual ARQ protocol, making the comparison conservative with respect to
FLUID.

To make a direct comparison between FLUID and ARQ, we align their source
packet streams at the block level. The FLUID source stream is partitioned
into blocks of $K$ source packets. For each FLUID block with block budget
\[
N = \left\lceil \lambda \cdot K \right\rceil,
\]
we compare against an ARQ block of $N$ source packets. Thus, FLUID delivers
$K$ source packets using budget $N$, while ARQ delivers $N$ source packets.

For analytical purposes, the FLUID execution is extended in the comparison
protocol model so that transmission continues until feedback indicates that
the receiver has obtained $N$ encoded packets, even though the block can be
recovered once $K$ encoded packets have been received. In the actual FLUID
protocol, the receiver can recover the block once it has received $K$ encoded
packets, and the sender stops once it receives feedback indicating recovery.
The analytical extension is used only to align the FLUID and ARQ transmission
processes, and is conservative with respect to FLUID.

Under this comparison protocol model and analytical extension, FLUID and ARQ
generate transmission events at exactly the same times and experience the
same loss realization for each transmission event. The packet contents differ:
ARQ retransmits source packets, whereas FLUID transmits encoded packets. The
difference between the protocols lies in recovery condition and delivery efficiency.
In FLUID, the receiver recovers the block once $K$ encoded packets have been received, whereas in ARQ, delivery finishes only after all $N$ source packets of the corresponding block have been received.
Thus, although the transmission processes are identical under the extended
model, FLUID finishes earlier.

This formulation enables a direct comparison of delivery latency, delivery efficiency,
and transmission overhead between the two protocols.

\subsection{Analytical Framework}

The analysis is performed at the block level. In practice, both FLUID and ARQ
deliver continuous streams of data rather than isolated blocks. FLUID forms
blocks explicitly, where each block contains $K$ source packets, with $K$
potentially varying from block to block. These blocks are formed from packets
arriving over a short time interval and often align with application-level
objects, such as video frames in streaming applications or windows of data in
bulk data delivery applications.

Although ARQ protocols do not explicitly define blocks, for analysis we
partition the ARQ packet stream into blocks aligned with the FLUID blocks.
This enables a direct comparison at the level of block delivery.

For a fixed aligned block, both FLUID and ARQ have the same causal structure
under the comparison protocol model. Packets are transmitted, losses are
reported through feedback, and additional packets are transmitted in response
to those reported losses. This creates a causal dependence between
transmissions: later transmissions for a block are triggered by feedback from
earlier transmissions for that block.

To capture this causal dependence, we analyze both protocols using a
round-based abstraction. Rounds are defined purely for analysis and are not
explicitly maintained by the protocols. In this abstraction, round 1 consists
of the initial $N$ packets transmitted for the block. For $r \ge 1$, round
$r+1$ consists of the packets transmitted in response to losses among packets
in round $r$. Each packet is assigned to the round whose feedback triggered
its transmission.

Because the underlying data stream is continuous, packets corresponding to
multiple blocks may be interleaved in time, so that packets for several blocks
may be in transit simultaneously. This interleaving reflects the operation of
the data stream and does not affect the analysis of an individual block.

The round assignment is causal rather than purely temporal. Rounds form
disjoint sets of packets, even though transmissions assigned to different
rounds may overlap in time in a continuous implementation. For analysis,
losses are attributed to the round in which the corresponding packets were
transmitted.

Under this abstraction, the number of lost packets evolves multiplicatively
across rounds. Round $r+1$ contains one packet transmitted in response to each
loss from round $r$, and those packets are themselves subject to loss. Thus,
the number of losses after each round decreases according to the product of
the loss fractions across rounds. This multiplicative evolution follows
directly from the causal dependence between rounds and does not depend on the
timing of transmissions.

When the block formation time is small relative to the RTT, each round of
transmissions for a block is completed in a short burst before feedback from
that round is received. In particular, the initial round of $N$ packets is
transmitted before feedback from these packets arrives; this feedback is
received after approximately one RTT and triggers the next round of
transmissions. The same pattern repeats for subsequent rounds. As a result,
for such blocks, rounds are sequential in time at the RTT time scale, with
each round separated from the next by approximately one RTT. The same timing
structure applies to ARQ under the aligned block decomposition.

Consequently, the number of rounds required to deliver a block determines
the delivery latency in units of RTT.

\subsection{Loss-Product Analysis}

We now characterize how packet losses evolve across rounds under the
analytical framework described above and derive the condition for delivery
to finish on this loss trajectory.

Let $N$ denote the number of packets transmitted in the initial round
for the block.

In round 1, $N$ packets are transmitted. Let $L_1$ denote the number of
packets lost in round 1. Define the loss fraction
\[
f_1 = \frac{L_1}{N},
\]
so that
\[
L_1 = f_1 \cdot N.
\]

For $r \ge 2$, round $r$ consists of $L_{r-1}$ packets. Let $L_r$ denote
the number of packets lost in round $r$, and define
\[
f_r = \frac{L_r}{L_{r-1}},
\]
so that
\[
L_r = f_r \cdot L_{r-1}.
\]

Thus,
\[
L_2 = f_2 \cdot L_1, \quad
L_3 = f_3 \cdot L_2, \quad \ldots, \quad
L_\ell = f_\ell \cdot L_{\ell-1}.
\]

By repeated substitution,
\[
L_\ell = \left(\prod_{r=1}^{\ell} f_r\right) \cdot N.
\]

Define the loss-product
\[
\pi_\ell = \prod_{r=1}^{\ell} f_r,
\]
so that
\[
L_\ell = \pi_\ell \cdot N.
\]

This multiplicative evolution follows directly from the round-based
structure and does not depend on the timing of transmissions.

After round $\ell$, the receiver has received $N - L_\ell$ packets.
The condition for delivery to finish can be expressed in terms of a slack
parameter $S$, defined as the maximum number of packet losses that can be
tolerated:
\[
L_\ell \le S.
\]

Using $L_\ell = \pi_\ell \cdot N$, this yields a general condition on
the loss-product.

\medskip
\noindent
\textbf{Loss-Product Rule.}
Delivery finishes by the first round $\ell$ for which
\[
\pi_\ell \le \frac{S}{N}.
\]

\medskip

We now specialize this rule to ARQ and FLUID.

For ARQ, all packets must be received, so $S = 0$. The condition becomes
\[
\pi_\ell \cdot N \le 0,
\]
which implies
\[
\pi_\ell = 0.
\]

\medskip
\noindent
\textbf{ARQ Loss-Product Condition.}
Delivery finishes by the first round $\ell$ for which
\[
\pi_\ell = 0.
\]

\medskip

For FLUID, the receiver can decode the block once it has obtained at least
$K$ encoded packets. Thus,
\[
S = N - K.
\]

The block budget is defined as
\[
N = \left\lceil \lambda \cdot K \right\rceil,
\]
and the slack satisfies
\[
S = N - K \ge \epsilon \cdot N,
\quad \text{where} \quad
\epsilon = \frac{\lambda - 1}{\lambda}.
\]

Thus,
\[
\pi_\ell \cdot N \le \epsilon \cdot N
\;\;\Rightarrow\;\;
\pi_\ell \le \epsilon.
\]
Since rounding may make $S/N$ slightly larger than $\epsilon$, the exact
slack threshold is $S/N$, while the condition $\pi_\ell \le \epsilon$
provides a simple conservative threshold.

\medskip
\noindent
\textbf{FLUID Loss-Product Condition.}
Delivery is guaranteed to finish by the first round $\ell$ for which
\[
\pi_\ell \le \epsilon.
\]

\medskip

The sequence
\[
L_\ell = \pi_\ell \cdot N
\]
defines the loss trajectory across rounds. This trajectory is common to
both FLUID and ARQ under the shared analytical framework; the protocols
differ in the slack threshold applied to the trajectory. For FLUID,
$S = N-K$, which is approximately $\epsilon \cdot N$ up to rounding,
whereas for ARQ, $S = 0$.

\begin{figure}[t]
\centering
\includegraphics[width=0.9\linewidth, trim=0cm 3cm 0cm 3cm, clip]{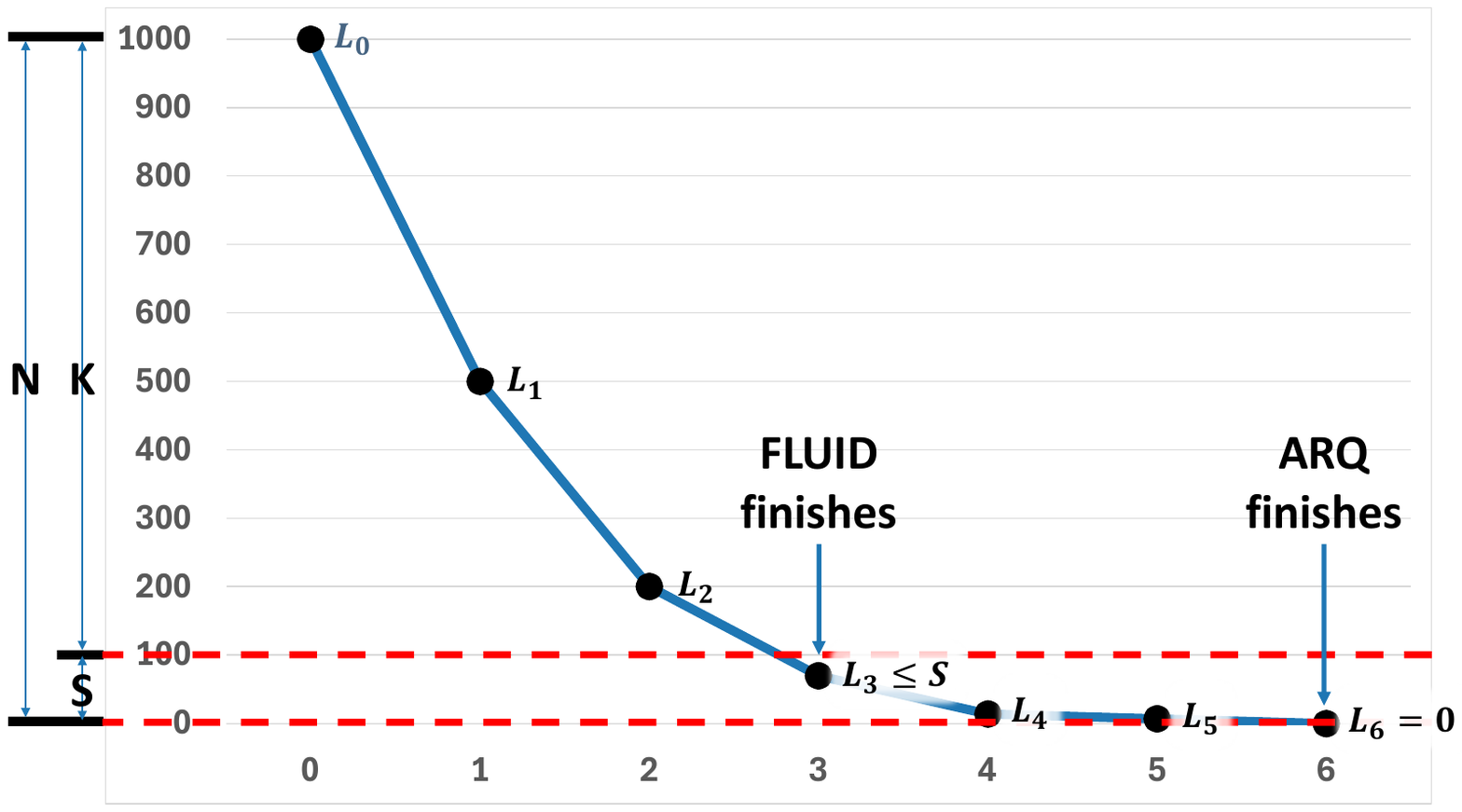}
\caption{
Loss-product trajectory $L_\ell = \pi_\ell \cdot N$ across rounds for the
aligned comparison with $N=1000$ and $\epsilon=0.10$. FLUID delivers
$K=900$ source packets with slack $S=N-K=100=\epsilon \cdot N$, while ARQ
delivers $N=1000$ source packets. The initial value is $L_0=N$, and each
round reduces the remaining losses multiplicatively according to the
realized loss fractions. The horizontal line at $S$ denotes the FLUID
slack. FLUID finishes delivery once $L_\ell \le S$, while ARQ continues
until $L_\ell = 0$.
}
\label{fig:loss_product_trajectory}
\end{figure}

Figure~\ref{fig:loss_product_trajectory} illustrates this common trajectory
for the aligned comparison with $N=1000$ and $\epsilon=0.10$. Thus, FLUID
delivers a block of $K=900$ source packets with slack
$S=N-K=100=\epsilon \cdot N$, while ARQ delivers a block of $N=1000$
source packets. FLUID finishes once the loss trajectory falls below the
slack threshold, while ARQ continues until the trajectory reaches zero.

The Loss-Product Rule characterizes how losses propagate across rounds.
For FLUID, delivery finishes once the loss-product falls below $\epsilon$.
For ARQ, zero slack requires the loss-product to be zero. Thus, FLUID
finishes delivery without requiring a zero-loss round, whereas ARQ depends
on one.

The Loss-Product Rule therefore provides a simple and intuitive way to
reason about FLUID delivery behavior under arbitrary network conditions.
In the following sections we use this framework to compare the number of
transmission rounds required by FLUID and ARQ under representative loss
scenarios.

\begin{figure}[ht]
\centering
\includegraphics[width=\linewidth, trim=2cm 4.6cm 2cm 6cm, clip]{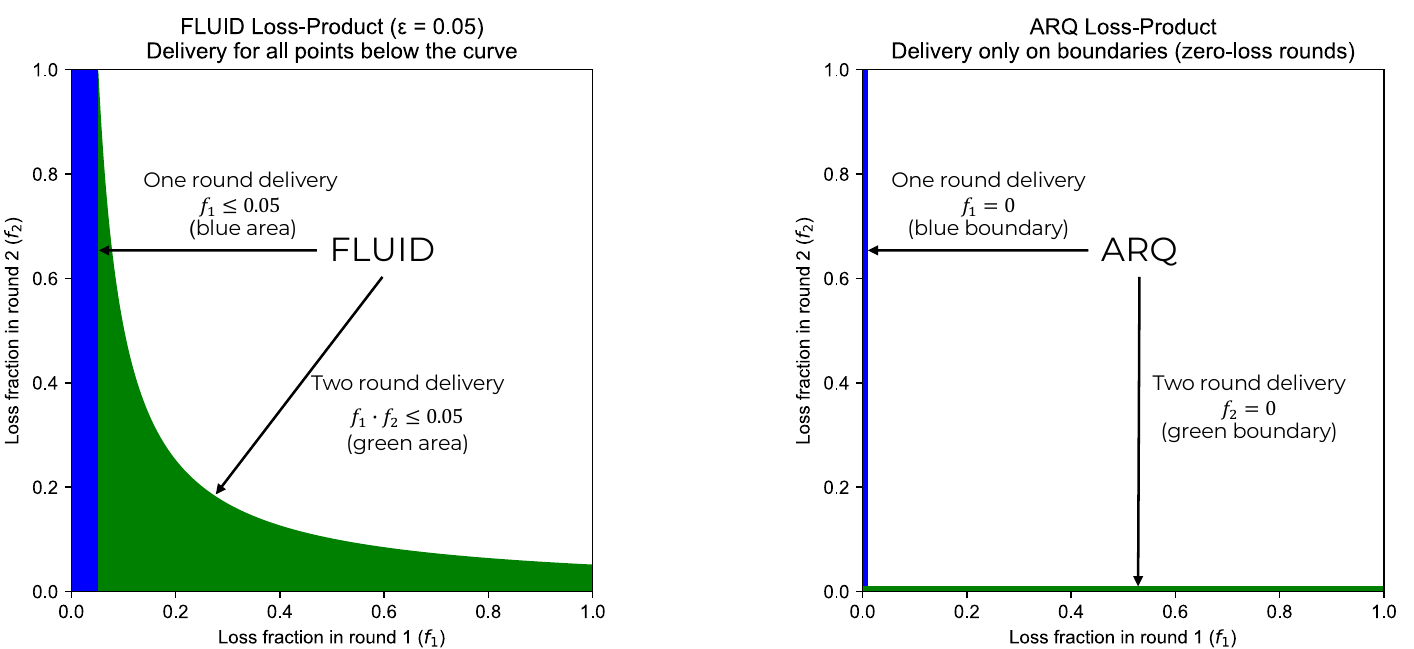}
\caption{
Comparison of FLUID and ARQ delivery conditions under packet loss for a
representative FLUID setting with $\epsilon=0.05$. The horizontal axis shows
the fraction of loss in round 1 ($f_1$) and the vertical axis shows the
fraction of loss in round 2 ($f_2$). For FLUID (left), delivery occurs for
all points below the curve defined by $f_1 \cdot f_2 \le 0.05$, illustrating
that delivery depends on the loss-product across rounds. For ARQ (right),
delivery occurs only on the boundaries where a round experiences zero loss
($f_1 = 0$ or $f_2 = 0$). Blue indicates delivery in one round and green
indicates delivery in two rounds.
}
\label{fig:arq_fluid}
\end{figure}

Figure~\ref{fig:arq_fluid} illustrates the one- and two-round delivery
regions for a representative FLUID setting with $\epsilon=0.05$. Later
numerical examples use $\epsilon=0.10$; these correspond to different
choices of slack on the same latency-efficiency tradeoff.

An important property of the loss-product formulation is that it does not
depend on any particular statistical model of packet loss. The rule applies
equally well to constant loss, burst loss, or highly uneven loss patterns.
Because the analysis depends only on the sequence of observed loss fractions
across rounds, it captures the progress of the protocol even when network
conditions vary significantly over time.

\FloatBarrier

\subsection{Delivery Latency Bound}

The round-based analysis also yields a bound on delivery latency. Let $T$ denote the time required to transmit the initial round of $N$ packets. The last packet of round 1 is transmitted by time $T$, and feedback for this packet is received after approximately one round-trip time (RTT), at time $T + \text{RTT}$.

Packets in subsequent rounds are triggered by feedback from the previous round. In the worst case, the last packet of round $\ell$ is triggered by the last feedback from round $\ell-1$. By induction, the last packet of round $\ell$ is transmitted no later than time $T + (\ell-1)\cdot \text{RTT}$, and the corresponding feedback is received by time $T + \ell \cdot \text{RTT}$.

It follows that if the block is delivered after $\ell$ rounds, where $\ell$ is the first round for which the Loss-Product Rule is satisfied, then the delivery latency for the block is at most
\[
T + \ell \cdot \text{RTT}.
\]
In the common case where the initial round transmission time $T$ is much smaller than RTT, this bound reduces to approximately $\ell \cdot \text{RTT}$.

More generally, this bound is conservative when transmissions from different rounds overlap in time. In particular, losses that occur early within a round can trigger subsequent transmissions well before the end of that round, so that later rounds may already be partially completed when earlier rounds finish. As a result, the actual delivery time may be smaller than the bound. In contrast, when transmissions occur in distinct bursts separated by feedback, the bound is relatively tight.

We also derive a simple (and generally loose) lower bound on delivery latency. The term $T$ reflects the time required to transmit the initial $N$ packets. Even if feedback is immediate and later rounds overlap in time, all packets in subsequent rounds are causally triggered by transmissions in earlier rounds, and thus the total number of packets that can have been received across all rounds cannot exceed the number of packets transmitted so far from round 1. Since delivery cannot finish until at least $K$ packets have been received, the delivery time must include at least the time required to transmit the first $K$ packets of round 1. Therefore, the delivery time must include at least the time required to transmit the first $K$ packets of round 1. Because $K$ is typically close to $N$, this lower bound is typically close to $T$.

Delivery latency therefore lies between a value close to $T$ and at most $T + \ell \cdot \text{RTT}$. Delivery latency is dominated by $T$ when $T$ is large relative to the RTT. In the limiting case where $\text{RTT} = 0$, lost packets are replaced immediately, and delivery latency reduces to the time required to transmit the first $K$ packets. In typical real-time data delivery settings where $T$ is much smaller than $\text{RTT}$, for example, $T = 1\,\text{ms}$ and $\text{RTT} = 50\,\text{ms}$, delivery latency is dominated by the $\ell \cdot \text{RTT}$ term determined by the Loss-Product Rule.

\subsection{Delivery Efficiency and Transmission Cost}

For ARQ, under the idealized model, all $N$ source packets are delivered to
the receiver, and thus the delivery efficiency is $1$.

For FLUID, for a block of $K$ source packets the sender has a budget of
\[
N = \left\lceil \lambda \cdot K \right\rceil
  = \left\lceil \frac{K}{1-\epsilon} \right\rceil
\]
encoded packets. Under the analytical extension used for comparison, FLUID is
run until the receiver has obtained $N$ encoded packets. Thus, in this
comparison, FLUID delivers $K$ source packets using $N$ received encoded
packets, and its delivery efficiency satisfies
\[
\frac{K}{N} \ge (1-\epsilon)\cdot \frac{K}{K+1},
\]
where the $\frac{K}{K+1}$ factor is due to rounding when $N$ is calculated
based on $K$. As $K$ grows, this lower bound on FLUID delivery efficiency approaches
$1-\epsilon$.

\medskip

The transmission cost follows from the same analytical extension. Under
the comparison protocol model, FLUID and ARQ generate the same transmission
events and experience the same loss realization for each event. Therefore, for
the corresponding FLUID and ARQ blocks, both protocols transmit the same total
number of packets.

Since FLUID delivers $K$ source packets while ARQ delivers $N$ source packets,
the ratio of FLUID transmissions per delivered source packet to ARQ
transmissions per delivered source packet is
\[
\frac{N}{K}.
\]
Using
\[
N = \left\lceil \frac{K}{1-\epsilon} \right\rceil,
\]
we have
\[
\frac{N}{K} \le \frac{1}{1-\epsilon} + \frac{1}{K},
\]
which approaches $\frac{1}{1-\epsilon}$ as $K$ grows.

\section{Comparison Under Random Loss}

This section gives an exact probabilistic comparison of FLUID and ARQ under
the comparison protocol model. In each transmission event, each packet is
independently lost with probability $p$ and received with probability $1-p$.

We use the aligned block comparison introduced above. ARQ delivers a block of
$N$ source packets. FLUID delivers a block of $K$ source packets using the same
block budget $N$. Throughout this section we fix
\[
N = 100.
\]
For FLUID, we use $\epsilon = 0.10$, so that $K = 90$ and the delivery efficiency is
$90\%$. For ARQ, all $N = 100$ source packets must be received.

The objective is to compute, for each protocol, the exact probability
distribution over the number of rounds required to finish delivery of a block.

\subsection{Distribution of Delivery Rounds}

Index the $N$ initial transmission positions by $1,\ldots,N$. For each
position, if the packet transmitted in one round is lost, the comparison
protocol model transmits one corresponding packet for that position in the
next round. For ARQ, this is a retransmission of the same source packet; for
FLUID, this is a newly generated encoded packet.

Let $X_\ell$ denote the number of these $N$ positions that have produced a
received packet after $\ell$ rounds. Under independent loss probability $p$
for each transmission event, a position remains unresolved after $\ell$
rounds only if all $\ell$ transmissions associated with that position are
lost. Thus each position is resolved with probability $1-p^\ell$, and
\[
X_\ell \sim \mathrm{Binomial}(N,\,1-p^\ell).
\]

For ARQ, delivery finishes when all $N$ packets have been received:
\[
T_{\mathrm{ARQ}} = \min \{ \ell : X_\ell = N \}.
\]

For FLUID, delivery finishes once at least $K$ encoded packets have been
received:
\[
T_{\mathrm{FLUID}} = \min \{ \ell : X_\ell \ge K \}.
\]

More generally, for a recovery threshold $M$, define
\[
T_M = \min \{ \ell : X_\ell \ge M \}.
\]
Then the probability that delivery finishes in round $\ell$ is
\[
\Pr(T_M = \ell)
=
\Pr(X_\ell \ge M) - \Pr(X_{\ell-1} \ge M),
\]
with $X_0 = 0$. The ARQ distribution is obtained by setting $M=N$, and the
FLUID distribution is obtained by setting $M=K$.

\begin{figure}[t]
\centering
\includegraphics[width=0.32\linewidth]{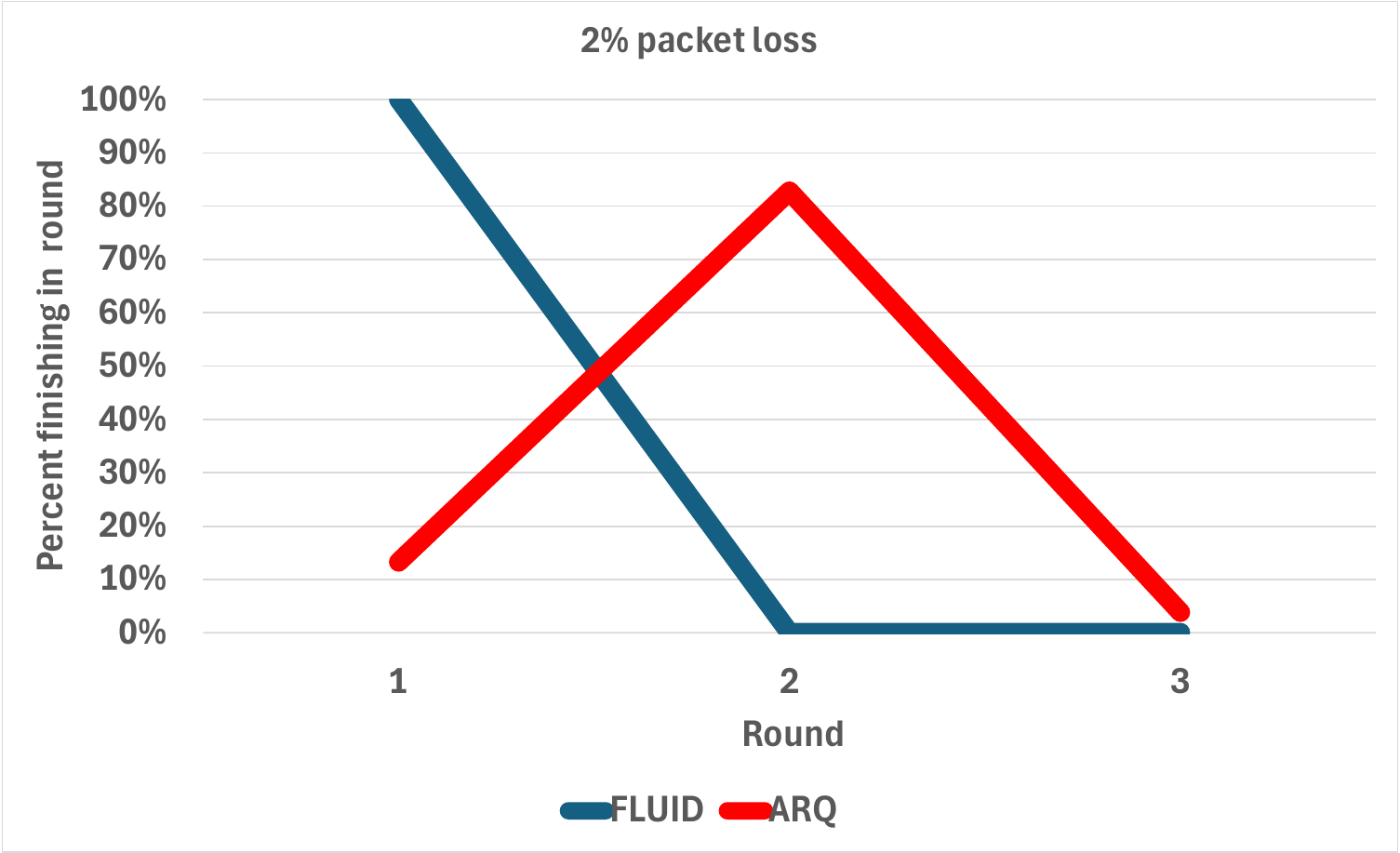}
\includegraphics[width=0.32\linewidth]{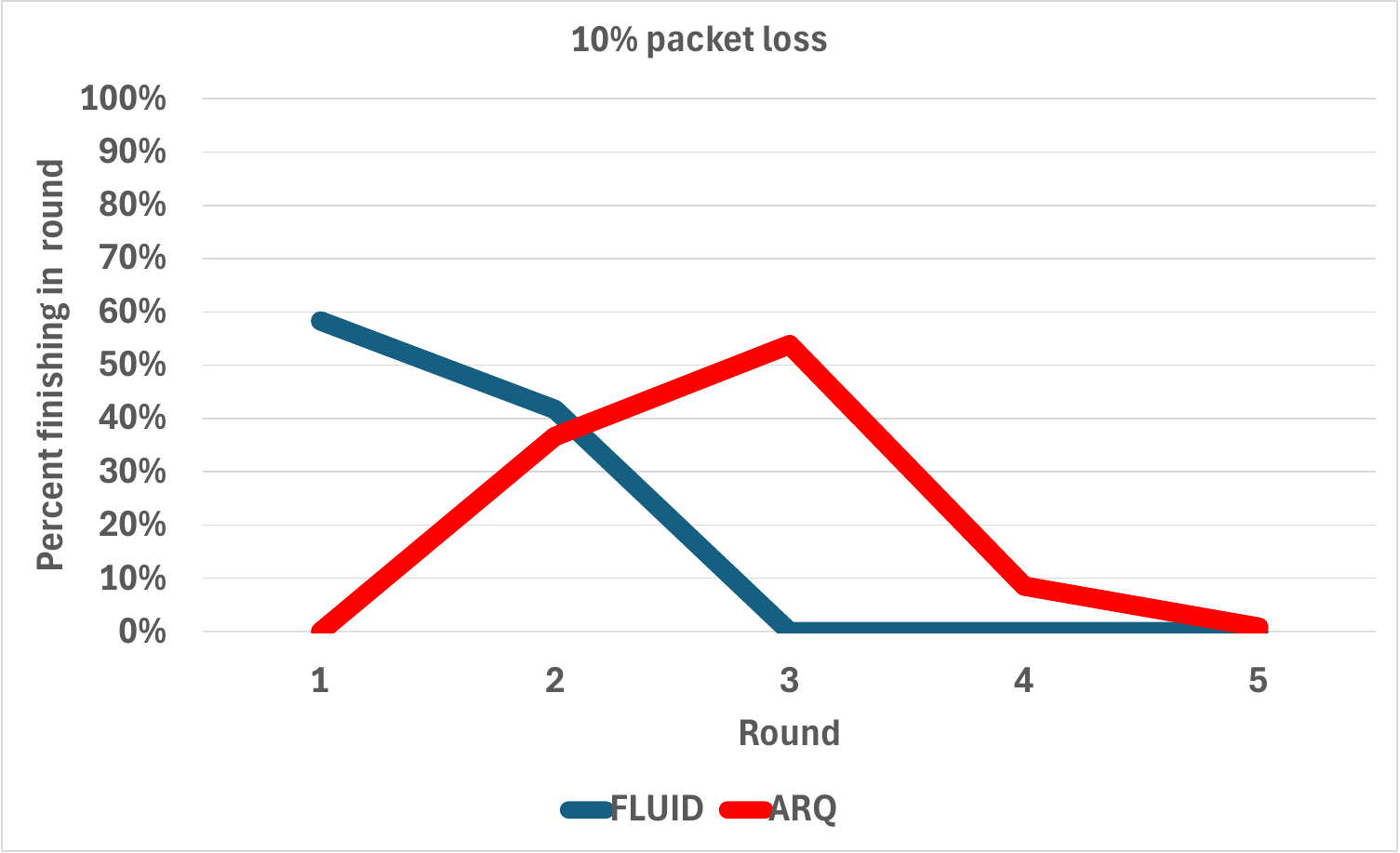}
\includegraphics[width=0.32\linewidth]{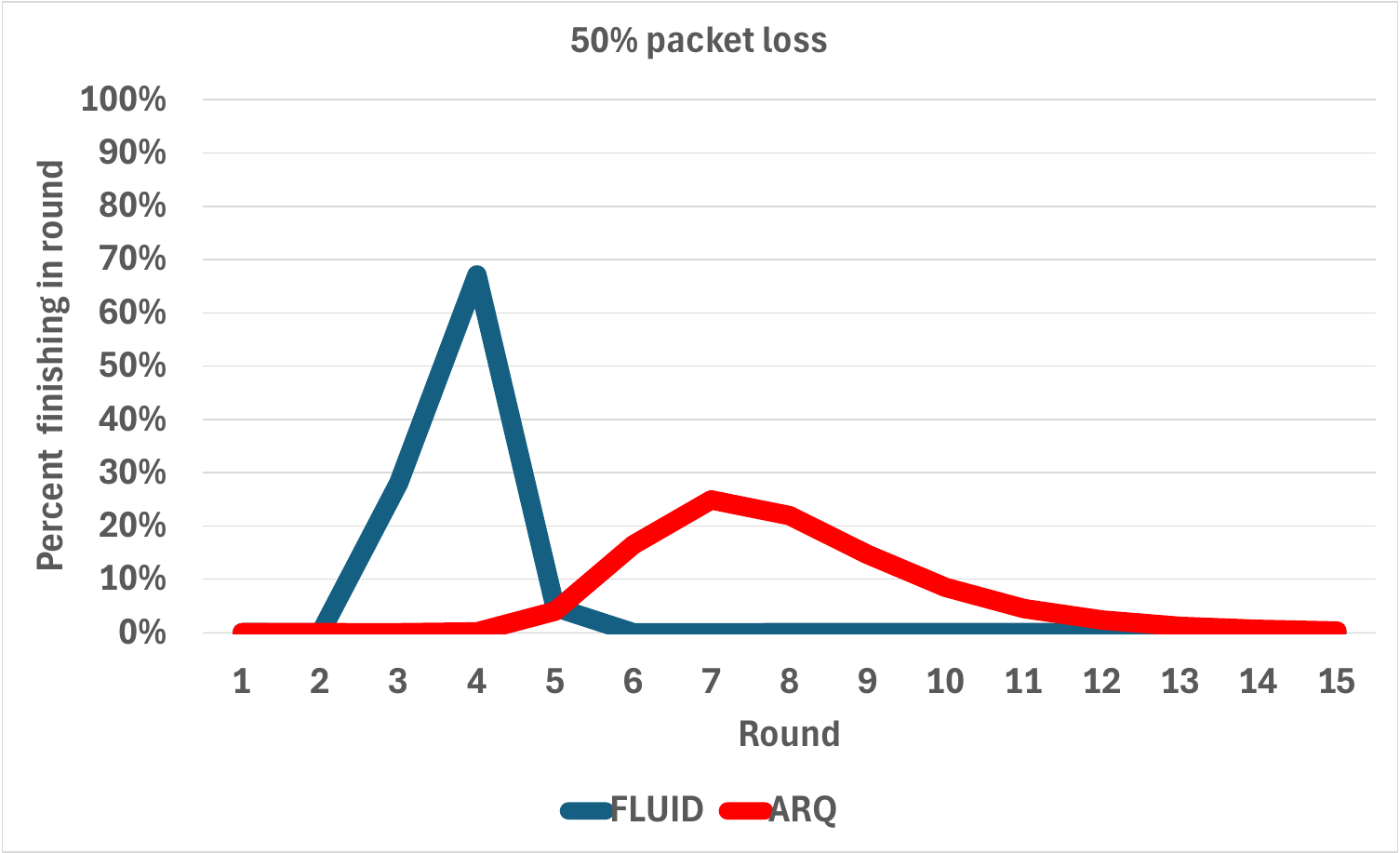}
\caption{
Percent finishing in each round under independent packet loss for the aligned comparison with $N=100$, shown for representative loss probabilities of $2\%$, $10\%$, and $50\%$. For FLUID, $\epsilon=0.10$, so $K=90$; for ARQ, the block contains $N=100$ source packets. FLUID concentrates finishing in early rounds, while ARQ spreads over more rounds and develops a heavier tail as loss increases.
}
\label{fig:round_distributions}
\end{figure}

Table~\ref{tab:round_distribution} shows the delivery-round distributions
for FLUID and ARQ in the aligned comparison with $N=100$, over packet loss
probabilities ranging from $0.1\%$ to $50\%$. For FLUID, $\epsilon=0.10$,
so $K=90$; for ARQ, the block contains $N=100$ source packets.
Figure~\ref{fig:round_distributions} illustrates the distribution of these delivery
rounds for representative loss probabilities of $2\%$, $10\%$, and $50\%$.

\begin{table}[t]
\centering
\small
\resizebox{\linewidth}{!}{
\begin{tabular}{|c|c|cccccccccc|}
\hline
 &  & \multicolumn{10}{c|}{Round number} \\
\hline
\% Loss & Scheme & 1 & 2 & 3 & 4 & 5 & 6 & 7 & 8 & 9 & 10+ \\
\hline

\multirow{2}{*}{0.1\%}
 & FLUID & 100.00 & -- & -- & -- & -- & -- & -- & -- & -- & -- \\
 & ARQ   & 90.48 & 9.51 & 0.01 & -- & -- & -- & -- & -- & -- & -- \\
\hline

\multirow{2}{*}{0.2\%}
 & FLUID & 100.00 & -- & -- & -- & -- & -- & -- & -- & -- & -- \\
 & ARQ   & 81.86 & 18.10 & 0.04 & -- & -- & -- & -- & -- & -- & -- \\
\hline

\multirow{2}{*}{0.5\%}
 & FLUID & 100.00 & -- & -- & -- & -- & -- & -- & -- & -- & -- \\
 & ARQ   & 60.58 & 39.17 & 0.25 & -- & -- & -- & -- & -- & -- & -- \\
\hline

\multirow{2}{*}{1.0\%}
 & FLUID & 100.00 & -- & -- & -- & -- & -- & -- & -- & -- & -- \\
 & ARQ   & 36.60 & 62.40 & 0.99 & 0.01 & -- & -- & -- & -- & -- & -- \\
\hline

\multirow{2}{*}{2.0\%}
 & FLUID & 100.00 & -- & -- & -- & -- & -- & -- & -- & -- & -- \\
 & ARQ   & 13.26 & 82.82 & 3.84 & 0.08 & -- & -- & -- & -- & -- & -- \\
\hline

\multirow{2}{*}{5.0\%}
 & FLUID & 98.85 & 1.15 & -- & -- & -- & -- & -- & -- & -- & -- \\
 & ARQ   & 0.59 & 77.26 & 20.90 & 1.18 & 0.06 & -- & -- & -- & -- & -- \\
\hline

\multirow{2}{*}{10.0\%}
 & FLUID & 58.32 & 41.68 & -- & -- & -- & -- & -- & -- & -- & -- \\
 & ARQ   & -- & 36.60 & 53.88 & 8.53 & 0.90 & 0.09 & 0.01 & -- & -- & -- \\
\hline

\multirow{2}{*}{20.0\%}
 & FLUID & 0.57 & 99.21 & 0.22 & -- & -- & -- & -- & -- & -- & -- \\
 & ARQ   & -- & 1.69 & 43.10 & 40.41 & 11.65 & 2.51 & 0.51 & 0.10 & 0.02 & -- \\
\hline

\multirow{2}{*}{50.0\%}
 & FLUID & -- & 0.01 & 28.09 & 67.10 & 4.77 & 0.03 & -- & -- & -- & -- \\
 & ARQ   & -- & -- & -- & 0.16 & 4.02 & 16.52 & 24.94 & 21.97 & 14.63 & 17.76 \\
\hline

\end{tabular}
}
\caption{
Percent finishing in round $\ell$ under independent packet loss for the aligned comparison with $N=100$. 
For FLUID, $\epsilon=0.10$, so $K=90$; for ARQ, the block contains $N=100$ source packets. 
The column labeled $10+$ aggregates all rounds $\ell \ge 10$, computed as $100\%$ minus the sum of rounds $1$ through $9$. 
Entries smaller than $0.01\%$ are omitted for readability.
}
\label{tab:round_distribution}
\end{table}

\FloatBarrier

\subsection{Interpretation}

The results highlight the tradeoff created by the FLUID slack. With
$\epsilon=0.10$, FLUID delivers $90$ source packets using a block budget of
$100$, while ARQ delivers all $100$ source packets. In exchange for this
$90\%$ delivery efficiency, FLUID finishes delivery in fewer rounds and with a much
tighter distribution.

This behavior is consistent across Table~\ref{tab:round_distribution}. At
$2\%$ loss, FLUID finishes essentially entirely in one round, while ARQ
finishes mostly in two rounds and has a tail out to four rounds. At $5\%$
loss, FLUID finishes mostly in round one and the remainder of the time in
round two, while ARQ mostly finishes in round two but with a significant
fraction of the time in round three and a tail into rounds four and five.
At $10\%$ loss, FLUID finishes within one to two rounds, while ARQ mostly
finishes in rounds two to four, with a tail into rounds five through seven.
At $20\%$ loss, FLUID finishes mostly in round two with a tail into round
three, while ARQ mostly finishes in rounds three to six, with a tail out to
round nine. At $50\%$ loss, FLUID mostly finishes in rounds three to five
with a small tail into round six, while ARQ exhibits a much broader
distribution, with most mass spread across rounds four through nine and a
substantial tail in round ten and beyond.

Across this range of loss probabilities, FLUID with slack parameter $\epsilon=0.10$,
corresponding to $90\%$ efficiency, yields a consistent reduction in
typical delivery latency by a factor of $2$ to $3$, as measured by where
the bulk of the distribution lies. A more
significant difference appears in the tail. ARQ finishing depends on rare
events involving the final remaining packets, which creates a persistent tail
in the distribution of delivery time. In contrast, FLUID finishes once a
fixed fraction of packets have been received, avoiding this final-packet
effect and yielding sharply concentrated finishing times. As a result, the
improvement in tail latency is substantially larger than the $2$ to $3$
factor observed for typical latency.

Even very small tail probabilities can have a noticeable impact in practice.
For example, at $120$ frames per second, a $0.1\%$ probability of requiring
an additional round corresponds to roughly one such event every eight
seconds on average. Over a sustained stream, these occasional late finishes
can introduce visible glitches in real-time applications such as streaming
video. By sharply reducing this tail, FLUID not only reduces latency but also
significantly improves consistency of delivery.

\section{Evaluation Under Real-World Streaming Workloads}

We evaluate FLUID using live video streaming workloads over real-world and controlled network conditions. The implementation is part of the LT3 system, a network-layer packet delivery system used in production environments. Tests are conducted using live streaming applications over cellular networks, including cross-continent streams. Each run corresponds to approximately 9--10 minutes of continuous streaming.

\subsection{Real-World Streaming Performance}

We evaluate FLUID using live video streaming workloads over real-world network conditions. The workload consists of HTTP-based live streaming (Twitch/Kick-like platforms) viewed over cellular connections, with streams originating from Europe and delivered to a client in the U.S. (Atlanta). Each run corresponds to approximately 9--10 minutes of continuous playback. The baseline corresponds to the same streaming workload over conventional transport on the identical network path, without coded delivery.

Table~\ref{tab:realworld} summarizes buffering behavior across multiple runs under these conditions.

\begin{table}[h]
\centering
\begin{tabular}{|c|ccc|ccc|}
\hline
Run & \multicolumn{3}{c|}{Buffer Events} & \multicolumn{3}{c|}{Buffer Time} \\
 & Baseline & FLUID & Reduction & Baseline & FLUID & Reduction \\
\hline
1 & 9 & 1 & 89\% & 5:58 & 0:04 & 99\% \\
2 & 5 & 0 & 100\% & 1:23 & 0:00 & 100\% \\
3 & 3 & 3 & 0\% & 1:12 & 0:10 & 86\% \\
4 & 5 & 2 & 60\% & 2:16 & 0:10 & 93\% \\
5 & 5 & 3 & 40\% & 0:54 & 0:10 & 81\% \\
6 & 7 & 6 & 14\% & 2:37 & 0:40 & 75\% \\
\hline
\end{tabular}
\caption{Buffering behavior over real-world live streaming runs. Each run is approximately 9--10 minutes.}
\label{tab:realworld}
\end{table}

Across all runs, FLUID reduces both the frequency and duration of buffering events. In several runs, buffering is eliminated entirely. When buffering occurs, events are shorter and recovery is faster.

\FloatBarrier

\subsection{Controlled Impairment Results}

To assess robustness, we evaluate under controlled impairment scenarios using repeatable network conditions. Each run uses the same impairment sequence, including bandwidth variation, latency variation, and periods of severe degradation such as high packet loss sustained over multiple seconds and complete outages (100\% loss) lasting several seconds. These conditions are designed to induce buffering and stress recovery behavior.

Table~\ref{tab:controlled} summarizes performance across these controlled scenarios.

\begin{table}[h]
\centering
\begin{tabular}{|c|ccc|ccc|}
\hline
Run & \multicolumn{3}{c|}{Buffer Events} & \multicolumn{3}{c|}{Buffer Time} \\
 & Baseline & FLUID & Reduction & Baseline & FLUID & Reduction \\
\hline
1 & 10 & 2 & 80\% & 0:57 & 0:28 & 50\% \\
2 & 8 & 4 & 50\% & 1:11 & 0:41 & 42\% \\
3 & Failure & 4 & N/A & 1:22 & 0:44 & 57\% \\
4 & Failure & 2 & N/A & 1:29 & 0:27 & 70\% \\
5 & Failure & 3 & N/A & 2:18 & 0:27 & 80\% \\
6 & 5 & 1 & 80\% & 0:47 & 0:26 & 49\% \\
7 & 5 & 3 & 40\% & 0:44 & 0:30 & 32\% \\
8 & 5 & 4 & 20\% & 1:03 & 0:48 & 40\% \\
9 & Failure & 2 & N/A & 7:09 & 0:30 & 93\% \\
10 & 6 & 2 & 66\% & 1:04 & 0:29 & 55\% \\
\hline
\end{tabular}
\caption{Performance under controlled impairment conditions. ``Failure'' indicates the baseline did not recover. Percentage reduction is undefined in these cases and marked as N/A.}
\label{tab:controlled}
\end{table}

Under these conditions, FLUID consistently reduces buffering and completes delivery even in cases where the baseline fails to recover within the test interval.

\FloatBarrier

\subsection{Discussion}

These observations are consistent with the loss-product behavior described earlier. In particular, FLUID avoids the dependence on complete packet recovery that produces long-tail latency in retransmission-based delivery. This leads to both reduced typical latency and tighter tail behavior in practical streaming workloads.

\section{Illustrative Delivery Examples}

The Loss-Product Rule provides an intuitive way to reason about delivery
behavior under arbitrary packet loss patterns. This section gives several
representative examples using the round-based analysis developed above.

Throughout these examples we use
\[
\epsilon = 0.10.
\]
Thus, for FLUID, the source block size is $K = (1-\epsilon)\cdot N
= 0.9\cdot N$, corresponding to $90\%$ delivery efficiency. Under the aligned
comparison, ARQ delivers a block of $N$ source packets and corresponds to
the special case $S=0$, or equivalently $\epsilon=0$.

Thus, in each example below, FLUID finishes once the loss-product falls
below $0.10$, while ARQ finishes only if the loss-product reaches zero.
Since all displayed loss fractions are positive, ARQ has not delivered by
the time FLUID finishes in these examples.

\subsection{Example 0: Immediate Delivery in the First Round}

Consider a case where the first round experiences modest loss,
\[
f_1 = 0.06.
\]
The loss-product after the first round is
\[
\pi_1 = f_1 = 0.06.
\]
Since
\[
\pi_1 \le \epsilon = 0.10,
\]
the FLUID block is delivered in the first round.

For ARQ, however, delivery in the first round would require
\[
\pi_1 = 0.
\]
Since $f_1 = 0.06 > 0$, ARQ has not delivered the block after the first
round.

This example illustrates that FLUID achieves single-round delivery when
loss is at most $\epsilon$.

\subsection{Example 1: Uneven Loss Across Two Rounds}

Consider the uneven loss pattern
\[
f_1 = 0.70, \qquad f_2 = 0.14.
\]
The loss-product evolves as
\[
\pi_1 = 0.70,
\]
\[
\pi_2 = 0.70 \cdot 0.14 = 0.098.
\]
Since
\[
\pi_2 \le \epsilon = 0.10,
\]
the FLUID block is delivered at the end of round 2.

For ARQ, the loss-product after two rounds is still positive:
\[
\pi_2 = 0.098 > 0.
\]
Thus ARQ has not delivered the block by round 2 and must wait for a later
round in which the remaining loss is eliminated.

This example shows that FLUID can deliver even when both rounds have loss
above $\epsilon$, because delivery depends on the product of the loss
fractions across rounds.

\subsection{Example 2: Constant Loss}

Consider a network where each round experiences a constant packet loss rate
\[
f_1 = f_2 = 0.30.
\]
The loss-product after two rounds is
\[
\pi_2 = 0.30 \cdot 0.30 = 0.09.
\]
Since
\[
\pi_2 \le \epsilon = 0.10,
\]
the FLUID block is delivered in the second round.

For ARQ, the loss-product after two rounds is
\[
\pi_2 = 0.09 > 0,
\]
so ARQ has not delivered the block by round 2.

This example shows that under constant loss, the loss-product decreases
geometrically across rounds until it falls below $\epsilon$.

\subsection{Example 3: Severe Uneven Loss}

Consider a bursty loss pattern typical of wireless or satellite networks
experiencing temporary fades:
\[
f_1 = 0.90, \qquad f_2 = 0.40, \qquad f_3 = 0.25.
\]
The loss-product evolves as
\[
\pi_1 = 0.90,
\]
\[
\pi_2 = 0.90 \cdot 0.40 = 0.36,
\]
\[
\pi_3 = 0.90 \cdot 0.40 \cdot 0.25 = 0.09.
\]
Since
\[
\pi_3 \le \epsilon = 0.10,
\]
the FLUID block is delivered at the end of round 3.

For ARQ, the loss-product after three rounds remains positive:
\[
\pi_3 = 0.09 > 0.
\]
Thus ARQ has not delivered the block by round 3.

This example shows that FLUID delivery depends only on the loss-product,
regardless of how loss is distributed across rounds.

\section{Conclusion}

The analysis shows that FLUID achieves substantially tighter delivery
latency than ARQ by changing the delivery condition for a block. Under the
Loss-Product Rule, FLUID finishes delivery once the loss-product across
rounds falls below the slack parameter $\epsilon$, while idealized ARQ
corresponds to zero slack and requires the remaining loss to be eliminated.
Thus FLUID can finish in a small number of rounds even when every round
experiences packet loss.

At the same time, FLUID remains deterministically close to the
bandwidth-optimal ARQ baseline. The slack parameter $\epsilon$ controls both
the delivery efficiency and the transmission cost ratio relative to ARQ:
larger slack enables tighter delivery latency, while smaller slack keeps
FLUID closer to the bandwidth-optimal baseline.

This behavior is enabled by the structure of the FLUID protocol. Because
fountain-encoded packets for a block are interchangeable, receiver feedback
can be count-based rather than identity-based. The sender does not need to
identify or retransmit particular missing packets; instead, reported losses
trigger new encoded packets for the block.

FLUID can be used ubiquitously, from benign networks with occasional
unpredictable losses to the most challenging environments, delivering tighter
latency while maintaining high delivery efficiency and bounded transmission
cost.

\section*{Acknowledgments}

This material is based upon work supported by the National Science Foundation under Grant No.~2212574. The author is grateful to John Byers for detailed comments that greatly improved the presentation.

\bibliographystyle{plain}
\bibliography{references}

@misc{rfc793,
  author = {Postel, Jon},
  title = {Transmission Control Protocol},
  howpublished = {RFC 793},
  year = {1981}
}

@article{AggarwalLubyMinder2025Immersive,
  author  = {Pooja Aggarwal and Michael Luby and Lorenz Minder},
  title   = {Enabling Immersive Experiences in Challenging Network Conditions},
  journal = {arXiv preprint arXiv:2304.03732v2},
  year    = {2025},
  doi     = {10.48550/arXiv.2304.03732},
  url     = {https://arxiv.org/abs/2304.03732v2}
}

@inproceedings{luby2002lt,
  author = {Luby, Michael},
  title = {{LT} Codes},
  booktitle = {Proceedings of the 43rd Annual IEEE Symposium on Foundations of Computer Science (FOCS)},
  year = {2002}
}

@article{shokrollahi2009raptor,
  author = {Shokrollahi, Amin and Luby, Michael},
  title = {Raptor Codes},
  journal = {Foundations and Trends in Communications and Information Theory},
  volume = {6},
  number = {3-4},
  pages = {213--322},
  year = {2009}
}

@misc{rfc6330,
  author = {Luby, Michael and Shokrollahi, Amin and Watson, Mark and Stockhammer, Thomas},
  title = {RaptorQ Forward Error Correction Scheme for Object Delivery},
  howpublished = {RFC 6330},
  year = {2011}
}

@inproceedings{sundararajan2011network,
  title = {Network Coding Meets TCP: Theory and Implementation},
  author = {Sundararajan, Jay Kumar and Shah, Devavrat and Medard, Muriel and Mitzenmacher, Michael and Barros, Joao},
  booktitle = {Proceedings of ACM SIGCOMM},
  year = {2011}
}

@inproceedings{lacan2012tetrys,
  author    = {Thai, Tuan Tran and Lochin, Emmanuel and Lacan, J{\'e}r{\^o}me},
  title     = {Online Multipath Convolutional Coding for Real-Time Transmission},
  booktitle = {Proceedings of the 19th International Packet Video Workshop},
  address   = {Munich, Germany},
  month     = may,
  year      = {2012}
}

@misc{rfc5109,
  author = {Li, A. and others},
  title = {RTP Payload Format for Generic Forward Error Correction},
  howpublished = {RFC 5109},
  year = {2007}
}

@misc{michel2019quicfec,
  title = {Forward Erasure Correction for QUIC},
  author = {Michel, François and others},
  year = {2019},
  note = {IETF Draft / Research Proposals}
}

@inproceedings{chu2023strack,
  title = {S-Track: Low-Latency Network Data Transfer},
  author = {Chu, David and others},
  booktitle = {USENIX NSDI},
  year = {2023}
}

@inproceedings{poularakis2020multipath,
  title = {Coded Multipath Transport},
  author = {Poularakis, Konstantinos and others},
  booktitle = {IEEE INFOCOM},
  year = {2020}
}

@inproceedings{jacobson1988congestion,
title={Congestion Avoidance and Control},
author={Jacobson, Van},
booktitle={ACM SIGCOMM},
year={1988}
}

@book{bertsekas1992data,
title={Data Networks},
author={Bertsekas, Dimitri and Gallager, Robert},
year={1992},
publisher={Prentice-Hall}
}

@misc{luby2024fluid,
  author = {Luby, Michael and Minder, Lorenz},
  title = {Methods for Reliable Low Latency Data Delivery Using Erasure Codes and Feedback},
  year = {2024},
  note = {US Patent No. 11,863,317, issued Jan. 2, 2024}
}

@misc{rfc2018,
  author       = {Mathis, M. and Mahdavi, J. and Floyd, S. and Romanow, A.},
  title        = {{TCP Selective Acknowledgment Options}},
  howpublished = {RFC 2018},
  year         = {1996},
  month        = oct,
  doi          = {10.17487/RFC2018},
  note         = {IETF}
}

@misc{rfc6298,
  author       = {Paxson, V. and Allman, M. and Chu, J. and Sargent, M.},
  title        = {{Computing TCP's Retransmission Timer}},
  howpublished = {RFC 6298},
  year         = {2011},
  month        = jun,
  doi          = {10.17487/RFC6298},
  note         = {IETF}
}

@misc{rfc8985,
  author       = {Cheng, Y. and Cardwell, N. and Dukkipati, N. and Jha, P.},
  title        = {{The RACK-TLP Loss Detection Algorithm for TCP}},
  howpublished = {RFC 8985},
  year         = {2021},
  month        = feb,
  doi          = {10.17487/RFC8985},
  note         = {IETF}
}

@misc{rfc6363,
  author       = {Watson, M. and Begen, A. and Roca, V.},
  title        = {{Forward Error Correction (FEC) Framework}},
  howpublished = {RFC 6363},
  year         = {2011},
  month        = oct,
  doi          = {10.17487/RFC6363},
  note         = {IETF}
}

\end{document}